\documentclass{emulateapj}
\usepackage{amstext,amsmath,natbib}

\shorttitle{LMXBs in Early-Type Galaxies}
\shortauthors{JUETT}
\slugcomment{Accepted for publication in the Astrophysical Journal Letters}

\begin{document}
\newcommand{\Ch}{{\em Chandra}}

\title{On the Nature of X-ray Sources in Early-Type Galaxies}

\author{Adrienne~M.~Juett}

\affil{\footnotesize Department of Astronomy, University
of Virginia, Charlottesville, VA 22903;\\ ajuett@virginia.edu}

\begin{abstract}
We show that the observed relationship between the fraction of
low-mass X-ray binaries (LMXBs) found in globular clusters (GCs) and
the GC-specific frequency for early-type galaxies is consistent with
an LMXB formation model in which the field population of LMXBs is
formed {\em in situ} via primordial binary formation.  The suggestion
that a significant fraction of the field LMXB population in early-type
galaxies was formed in GCs is not required by the data.  Finally, we
discuss observational studies that will test this model more
thoroughly.
\end{abstract}

\keywords{binaries: close --- X-rays: binaries --- X-rays: galaxies
--- galaxies: elliptical and lenticular --- globular clusters:
general}

\section{Introduction}
Low angular resolution studies of early-type galaxies found that the
X-ray emission from these systems typically consists of two
components, a soft, thermal component from hot gas, and a harder,
power-law component attributed to unresolved X-ray binaries
\citep[e.g.,][]{mka+97}.  The superb angular resolution of the {\em
Chandra X-ray Observatory} has allowed for a significant fraction of
the power-law component to be resolved into point sources, confirming
the binary nature of the hard component \citep[e.g.,][]{sib00}.
Early-type galaxies are generally dominated by stellar populations
older than 1~Gyr.  Thus, the majority of X-ray binaries in these
galaxies are low-mass X-ray binaries (LMXBs), rather than the
short-lived ($\sim10^{7}$~yr) high-mass variety.  LMXBs consist of a
neutron star (NS) or black hole (BH) accreting material from a
$\lesssim 1 \, M_{\odot}$ donor.

A number of early-type galaxies have been studied with \Ch\/ in an
effort to understand the LMXB population of these galaxies.  Combined
X-ray and optical studies have found that 20\%--70\% of the LMXBs in
early-type galaxies are associated with globular clusters
\citep[GCs;][]{alm01,bsi01,sib01,ski+03,kmz02,kmz+03,mkz03,ssi03,hb04,jcf+04,rsi04}.
This is significantly larger than the roughly 10\% of LMXBs found in
GCs in the Milky Way.  \citet{wsk02}, using {\em ASCA} data, found
that the X-ray to optical luminosity ratio, $L_{\rm X}/L_{\rm opt}$,
was proportional to the GC-specific frequency, $S_N$ (the number of
GCs per unit optical luminosity), of the galaxies.  They found no
correlation between the stellar age of the galaxies and $L_{\rm
X}/L_{\rm opt}$.  \citet{wsk02} concluded that most or all of the
LMXBs in early-type galaxies were formed in GCs and further suggested
that the LMXBs not associated with GCs either had been ejected from
GCs or were formed in GCs that had since disrupted.  The similarities
in the properties of the GC LMXBs and those found in the fields of
early-type galaxies has been used as support for this scenario
\citep[e.g.,][]{mkz03}.

However, it has been noted that the fraction of LMXBs found in GCs
shows some dependence on $S_N$ \citep{mkz03,ski+03}.  In this Letter,
we show that the relationship between the fraction of LMXBs found in
GCs and $S_N$ is consistent with the field population of LMXBs being
formed via primordial binary formation in the fields of the galaxies.

\section{Model of LMXB Formation}
We assume that LMXB formation in early-type galaxies is the same as
commonly accepted for the Milky Way.  In this scenario, the field
LMXBs are formed {\em in situ} via primordial binary formation and the
LMXBs associated with GCs are formed through dynamical interactions in
the clusters.  For any galaxy, the fraction of LMXBs found in GCs,
denoted by $f_{\rm LMXB,GC}$, is equal to
\begin{equation}
f_{\rm LMXB,GC} = \frac{N_{\rm LMXB,GC}}{N_{\rm LMXB,GC} + N_{\rm
LMXB,F}} \, ,
\label{eqn:1}
\end{equation}
where $N_{\rm LMXB,GC}$ and $N_{\rm LMXB,F}$ are the numbers of LMXBs
found in GCs and the field, respectively.

It has long been recognized that GCs produce $\sim 100$ times more
LMXBs per unit mass than the field of the Galaxy.  This enhanced
formation efficiency is attributed to dynamical interactions between
NSs and stars in the dense environments of GCs \citep[e.g.,][]{c75}.
Observational evidence suggests that $\approx4$\% of GCs have a bright
($L_{\rm X} \gtrsim 5 \times 10^{37}$~erg~s$^{-1}$) LMXB regardless of
the host galaxy morphological type \citep[e.g.,][]{kmz02}.  Given the
similarities in the average properties of GCs in all galaxies
\citep[e.g.,][]{az98}, this constant LMXB formation efficiency is not
surprising and is a measure of the efficiency of dynamical formation
mechanisms.  We then expect that $N_{\rm LMXB,GC}$ is proportional to
the number of GCs in the galaxy.  The number of GCs is equal to $S_N$
times the galaxy luminosity in the appropriate units.  For our
purposes, it is more interesting to give $N_{\rm LMXB,GC}$ in terms of
the galaxy mass, so we assume a fixed mass-to-light ratio for our
sample galaxies.  Therefore, we find that
\begin{equation}
N_{\rm LMXB,GC} \propto S_N M_{\star} \, ,
\end{equation}
where $M_{\star}$ is the stellar mass of the galaxy.  Here, and
throughout, we neglect differences in the stellar mass-to-light ratio
of our sample galaxies as we are primarily concerned with early-type
systems.  Were we to compare early- and late-type galaxies, it would
be more appropriate to use the mass normalized GC-specific frequency
\citep[e.g.,][]{az98}.

LMXBs in the field of the Galaxy are the descendants of primordial
binaries containing a massive star.  LMXBs have a long ($\sim 1$~Gyr)
lag time between star formation and the X-ray active phase
\citep{wg98} and therefore track the star formation history over long
timescales.  For a single starburst model for star formation in
early-type galaxies, the number of LMXBs formed from primordial
binaries will be a function of the stellar mass involved in the star
formation event and the time since star formation occurred.
Initially, the number of active LMXBs will increase until $\sim 1$~Gyr
and then begin to decrease with time \citep{wg98}.  The exact time
dependence of the LMXB population is unknown.  In addition, complex
star-formation histories, such as small starbursts associated with
mergers, will complicate the time dependence for any galaxy.  For our
model, we assume that the time dependent variation of the field LMXB
population is small when comparing different early-type galaxies.
Under this assumption, $N_{\rm LMXB,F}$ is simply proportional to
$M_{\star}$.

\begin{figure}
\epsscale{1.2}
\centerline{\plotone{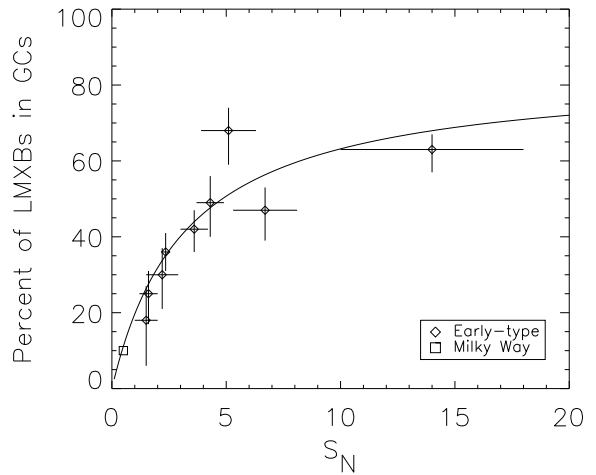}}
\caption{Fraction of the LMXB population found in GCs (given in
percent) plotted versus the GC-specific frequency, $S_N$ for 9
early-type galaxies.  The best-fit model is overplotted.  Included is
a data point for the Milky Way LMXB population which was not used in
the fit.}
\label{fig:1} 
\end{figure}

Plugging the relationships for $N_{\rm LMXB,GC}$ and $N_{\rm LMXB,F}$
into Equation~\ref{eqn:1} and rearranging, we find
\begin{equation}
f_{\rm LMXB,GC} = \left( 1 + \frac{C}{S_N} \right)^{-1} \, ,
\label{eqn:2}
\end{equation}
where $C$ is a constant that accounts for the formation efficiencies
in GCs and the field, and the mass-to-light ratio of the galaxies.  To
test this model, we have accumulated data from the literature for nine
early-type galaxies with both \Ch\/ observations and GC
identifications.  Table~\ref{tab:1} gives the values of $S_N$ and the
calculated fraction of LMXBs found in GCs.  We then fit these data
with a function of the form
\begin{equation}
f_{\rm LMXB,GC} = \left( A + \frac{B}{S_N} \right)^{-1} \, ,
\label{eqn:3}
\end{equation}
where $A$ and $B$ were both allowed to vary.  We find that $A =
1.19\pm0.19$ and $B = 3.9\pm1.0$, with a reduced $\chi^2$ for the fit
of 0.74.  The data and best-fit function are shown in
Figure~\ref{fig:1}.  For reference, we have included a data point for
the Milky Way with $S_N = 0.5$ and $f_{\rm LMXB,GC} \approx 0.1$
\citep{h91,lvv+01}.  The Milky Way data point includes both the disk
and bulge contributions.  As we see, the data follow the expected
trend from our model.  Therefore, we conclude that the field
population of LMXBs in early-type galaxies was formed in the field
from primordial binaries.  While it is intriguing that the best-fit
model also fits the Milky Way data point, it should be noted that for
a luminosity limit similar to those found in studies of early-type
galaxies, the fraction of LMXBs in GCs in the Milky Way is highly
uncertain because of low number statistics, difficulties in
determining the distance to Galactic LMXBs, and the issue of including
transient systems in the statistics.  As noted earlier, mass-to-light
ratio differences may also be important when comparing the GC-specific
frequency across morphological types.

We can also compare the prediction of this model to other observed
properties of the LMXBs in early-type galaxies.  \citet{wsk02} found
that $L_{\rm X}/L_{\rm opt}$ for the hard spectral component in
early-type galaxies was proportional to $S_N^{1.2\pm0.4}$.  Assuming
that the total X-ray luminosity of the LMXBs can be approximated by
$L_{\rm X,total} = \langle L_{\rm X}\rangle N_{\rm LMXB}$, where
$\langle L_{\rm X}\rangle$ is an average X-ray luminosity per binary,
assuming the same average luminosity for GC and field systems
\citep[see][]{ski+03}, and $N_{\rm LMXB}$ is the total number of LMXBs
in the galaxy, our model would predict that $L_{\rm X,total}/L_{\rm
opt}$ varies linearly with $S_N$.  While not the same relationship
found by \citet{wsk02}, we find that the data are equally well fitted
by our model.  It has also been shown that $N_{\rm LMXB}$ and $L_{\rm
X,total}$ in early-type galaxies are proportional to the stellar mass
of the galaxies with a weak dependence on the morphological type
\citep{g04}.  Our model would also predict that $N_{\rm LMXB}$ and
$L_{\rm X,total}$ are proportional to the stellar mass of the galaxies
with some dependence on the GC-specific frequency that could yield the
morphological type difference seen.  It should be noted that
\citet{g04} suggest that statistics and inaccurate calibration of the
near-infrared mass-to-light ratio may cause some of the morphological
dependence seen in their data.  Our model would predict that the
dependence on $S_N$ would be of the order of a factor of 1.5 given the
galaxies in the \citet{g04} sample, similar to what was measured.
However, the inclusion of late-type galaxies may actually decrease the
significance of this effect given the expected difference in the field
LMXB population formation.  A further study encompassing early-type
galaxies with a larger range of $S_N$ values is warranted.

The high ($L_{\rm X} \gtrsim 5 \times 10^{37}$~erg~s$^{-1}$) observed
luminosities of the LMXBs in early-type galaxies has been used to
suggest that such systems could not have formed from primordial
binaries since the X-ray lifetimes of such systems would be much
shorter than the stellar age of the galaxy.  The general argument is
that at $5 \times 10^{37}$~erg~s$^{-1}$ and with a $1 \, M_{\odot}$
companion, the system would have an X-ray active phase of less than
$\approx 100$~Myr.  There are many difficulties with this conclusion.
First, the lag time between binary formation and LMXB activity can be
quite long.  If the companion is a $1 \, M_{\odot}$ donor that starts
mass transfer after the donor has turned off the main sequence, then
the binary was formed more than 10~Gyr ago.  For systems that start
mass transfer on the main sequence, the time between the formation of
the binary and the start of mass transfer is dictated by either
gravitational radiation or magnetic braking, both of which have
characteristic timescales of the order of a few gigayears
\citep[e.g.,][]{prp03}.

It is also important to recognize that the mass-transfer rates
observationally inferred may not accurately reflect the long-term
mass-transfer rates due to short-timescale fluctuations, such as the
disk ionization instability or X-ray irradiation of the companion
\citep[e.g.,][]{v96,br04}.  As a result of such effects, intrinsically
low ($\sim 10^{-10} \, M_{\odot}$~yr$^{-1}$) mass-transfer rate
systems can masquerade as high mass-transfer rate systems.  All of the
BH LMXBs in the Milky Way, as well as a comparable number of NS
systems, exhibit X-ray outburst phases in which the maximum luminosity
is a significant fraction of the Eddington limit for the compact
object.  These systems, commonly referred to as soft X-ray transients
(SXTs), are thought to have low mass-transfer rates from the donor
such that the accretion disk is unstable to the disk ionization
instability \citep{v96}.  In SXTs, the systems undergo outbursts
lasting days to months (and even years in some cases) and then return
to a quiescent state.  The recurrence time can be as short as a year
or longer than the history of X-ray astronomy.  While the disk
instability model works well in predicting which systems are likely to
undergo this phenomena, it provides no clues as to the length of the
X-ray outbursts or recurrence times.  Other transient phenomena, such
the effect of X-ray irradiation on the donor \citep[e.g.,][]{br04},
can also cause intrinsically low mass-transfer systems to have
observed luminosities much higher than expected.

\section{Globular Cluster Contribution to Field LMXBs}
It has been suggested that GCs may contribute to the field LMXB
population through ejection of LMXBs or disruption of the clusters
\citep[e.g.,][]{wsk02}.  These alternatives have been invoked for a
number of reasons including the expected short X-ray active lifetimes
of LMXBs formed from primordial binaries and the similarities in the
X-ray properties of the field and GC populations of LMXBs.  But it is
important to recognize that once LMXBs leave the GC environments,
through either ejection or disruption, they will evolve like
primordial binaries and would have similar X-ray active lifetimes.
Therefore, GC formation of field LMXB scenarios require a constant
input of LMXBs into the field to account for the observed population.

The same dynamical processes that form LMXBs in GCs would be
responsible for the ejection of these systems.  Therefore, the number
of LMXBs ejected from GCs in any galaxy would be proportional to the
number of GCs or $S_N M_{\star}$.  If all of the field LMXBs were due
to ejections, we would expect no dependence of $f_{\rm LMXB,GC}$ on
$S_N$.  Our data rule out this scenario at greater than the 99.99\%
confidence level.  While not required, we can not rule out that some
small fraction of the field LMXB population was ejected from GCs with
the current data.  The upper limit on the fraction of the field
population due to ejections is dependent on $S_N$.  For $S_N=2.0$, we
find that $\lesssim20$\% of the field population of LMXBs could have
been ejected from GCs.

Theoretical studies of GC disruption find that a significant fraction
of the GCs in a galaxy will be disrupted with the exact fraction
showing some dependence on the galaxy properties
\citep[e.g.,][]{v00b,fz01}.  If we assume that disruption of a GC does
not change the probability for LMXB formation in the cluster, then to
fully account for the number of LMXBs seen in the field of early-type
galaxies, the number of disrupted clusters must be between 0.5 and 4
times the number of GCs currently observed.  Such numbers push the
limits of what is expected from current models for GC disruption
\citep{v00b}, but LMXBs from disruptions cannot be ruled out.  One
difficulty however, is that a significant fraction of the disruptions
may occur early in the history of the galaxy, in the first $\lesssim
3$~Gyr.  At early times, the dynamical interactions in GCs may be less
efficient at producing LMXBs \citep[e.g.,][]{dh98}.

\section{Discussion}
We have shown that the observed relationship between the fraction of
LMXBs found in GCs and the GC-specific frequency in early-type
galaxies is consistent with the field population of LMXBs forming {\em
in situ} via primordial binary formation.  To test our model, better
data for both $f_{\rm LMXB,GC}$ and $S_N$ are required.  In current
studies, only a fraction of the LMXB population in early-type galaxies
can be compared to the GC population because of differences in the
field of views of the instruments.  More thorough coverage of the
\Ch\/ field of view by GC studies will improve the statistics for
$f_{\rm LMXB,GC}$.  The values for $S_N$ are also subject to
observational uncertainties because of uncertainties in the estimates
of both the number of GCs and the optical magnitude of the galaxies.
Of particular note is the recent revision in $S_N$ for NGC 1399
\citep[see][]{drg+03}.

\begin{deluxetable*}{lccccc}
\tabletypesize{\footnotesize} 
\tablewidth{0pt} 
\setlength{\tabcolsep}{0.1in}
\tablecaption{Globular Cluster and LMXB Connection Data}
\tablehead{\colhead{Galaxy} & \colhead{$S_N$} & 
  \colhead{$N_{\rm LMXB,GC}$} & \colhead{$N_{\rm LMXB}$\tablenotemark{a}} & 
  \colhead{$f_{\rm LMXB,GC}$ (\%)} & \colhead{Refs\tablenotemark{b}}}
\startdata
NGC 1332 & 2.2$\pm$0.7 & 9 & 30 & $30^{+7}_{-9}$ & 1,2 \\
NGC 1399 & 5.1$\pm$1.2 & 26 & 38 & $68^{+6}_{-9}$ & 3,4 \\
NGC 1553 & 1.5$\pm$0.5 & 2 & 11 & $18^{+9}_{-12}$ & 5,6 \\
NGC 3115 & 1.6$\pm$0.4 & 9 & 36 & $25^{+6}_{-8}$ & 7,8 \\
NGC 4365 & 4.3$\pm$0.6 & 18 & 37 & $49^{+7}_{-9}$ & 5,9 \\
NGC 4472 & 3.6$\pm$0.6 & 30 & 72 & $42^{+5}_{-6}$ & 10,11 \\
NGC 4486 (M87) & 14$\pm$4 & 60 & 96 & $63^{+4}_{-6}$ & 12,13 \\
NGC 4649 & 6.7$\pm$1.4 & 22 & 47 & $47^{+6}_{-8}$ & 5,14 \\
NGC 4697 & 2.1$\pm$0.3 & 32 & 89 & 36$\pm$5 & 15,16 \\
\enddata
\label{tab:1}
\tablenotetext{a}{The total number of X-ray point sources covered by
the field of view of the optical globular cluster catalogs.}
\tablenotetext{b}{References: (1) \citet{kw01b}, (2) \citet{hb04}, (3)
\citet{drg+03}, (4) \citet{alm01}, (5) \citet{k97}, (6)
\citet{ski+03}, (7) \citet{pkt+04}, (8) \citet{kmz+03}, (9)
\citet{ssi03} (10) \citet{rz04}, (11) \citet{mkz03}, (12)
\citet{mhh94}, (13) \citet{jcf+04}, (14) \citet{rsi04}, (15) Jordan et
al. 2005, in prep., (16) Sivakoff et al. 2005, in prep.}
\end{deluxetable*}

A measurement of a time-dependence in the number of field LMXBs per
unit galaxy mass would also verify our model.  This measurement
requires good estimates of the stellar age of galaxies.  In addition,
understanding how both stellar age measurements and LMXB populations
are affected by complex star-formation histories is necessary to
interpret the results.  \citet{tfw+00} found that galactic age
estimates from single stellar population models are heavily weighted
by young stars.  No comparable study of the theoretical prediction for
LMXB populations has been performed.  As noted previously,
\citet{wsk02} found that the total X-ray luminosity attributed to
LMXBs in early-type galaxies was not correlated with stellar age, but
their analysis included the contributions from both field and GC
LMXBs.  The GC LMXBs are not expected to be dependent on the stellar
age of the galaxy, and therefore the inability to separate GC and
field LMXBs, due to the low spatial resolution of {\em ASCA}, may
explain why no age dependence was found in the \citet{wsk02} study.

The spatial distribution of GCs is generally more extended than the
galaxy light \citep[e.g.,][]{az98}.  In addition, GC spatial profiles
flatten at small radii.  Our model predicts that given the different
formation sites for GC and field LMXBs, there would be similar radial
distribution differences seen in the LMXB population.  \citet{kmz02}
found that the radial profiles of field and GC LMXBs were roughly
similar, although their study was limited in the number of systems (30
and 42 for GC and field LMXBs, respectively) and the radial extent
sampled.  Better coverage of the LMXB and GC populations is necessary,
especially at the inner and outer regions of the galaxies where the GC
and optical light profiles are the most different.  Two additional
effects must also be considered: {\em i}) how supernova kicks will
affect the distribution of primordial binaries in the field, and {\em
ii}) how the preference of GC LMXBs to be located in red, rather than
blue, GCs \citep{alm01,kmz02,kmz+03,ski+03,jcf+04} affects the GC LMXB
distribution.

Given the different formation mechanisms for field and GC LMXBs, it is
reasonable to assume that the distributions of the physical parameters
(i.e., orbital period, donor mass, and mass transfer rates) will be
different.  This, in turn, could yield differences in the observed
luminosity function.  Yet how different the luminosity function will
be is unknown.  At present, no theoretical study of the predicted
physical parameters of GC LMXBs exists.  Such a study is essential in
determining if a luminosity function difference is expected and/or
measurable.  Also, a better understanding of how transient behavior
affects the observed luminosity function is also necessary.
Multi-epoch observations of LMXBs in early-type galaxies will provide
more information on the transient behavior of LMXBs, although longer
observations are required to distinguish between real transient
behavior (luminosity changes of more than 3 orders of magnitude) and
the factor of $\sim 10$ variability commonly seen in persistent LMXBs
\citep[][and references therein]{lvv+01}.

Finally, it has been suggested that the lack of spectral differences
between field and GC LMXBs supports a common origin for these systems
\citep{mkz03}.  There is no observational evidence of differences in
the spectral parameters of field and GC LMXBs in the Milky Way
\citep[e.g.,][]{cs97}.  Therefore, it is unlikely that any spectral
differences would be seen, particularly given the low spectral
sensitivity of studies of LMXBs in early-type galaxies.

\acknowledgements{We acknowledge useful discussions with Eric Pfahl,
Saul Rappaport, and Gregory Sivakoff.  We would also like to thank
Craig Sarazin for comments on this Letter.  Support for this work was
provided by NASA through $Chandra$ Award Numbers GO3-4099X, AR3-4005X,
GO4-5093X, and AR4-5008X, issued by the $Chandra$ X-ray Observatory
Center, which is operated by the Smithsonian Astrophysical Observatory
for and on behalf of NASA under contract NAS8-39073. }

\end{document}